# Insights into neutron transport in a new multipurpose nuclear reactor

Luiz P. de Oliveira[1,*], Alexandre P.S. Souza[1,2], Carlos G.S. Santos[1,2],
Iberê R.S. Júnior[1], Barbara P.G. Silva[1], Marco A.S. Pereira[1], Frederico A. Genezini[2]

[1]Reator Multipropósito Brasileiro – RMB/CNEN, 2242 Prof. Lineu Prestes Av., São Paulo-SP, Brazil;
[2]Instituto de Pesquisas Energéticas e Nucleares – IPEN/CNEN, 2242 Prof. Lineu Prestes Av.,
São Paulo-SP, Brazil

## ABSTRACT

Neutron transport inside the reflector tank of the new Brazilian nuclear reactor, called the RMB (Brazilian Multipurpose Reactor), is investigated using the stochastic Monte Carlo method. One of the main characteristics of research nuclear reactors is the core power density, which is proportional to the neutron flux generated in the fission reaction. The RMB is part of the new generation of multipurpose reactors, and studying neutron transport is essential for the project to meet its objectives of producing radioisotopes, irradiating materials, and generating neutron beams for scientific investigations. The neutron moderation process in the reflector tank is very well described by Maxwellian distributions, which provide temperatures ranging from T = 331.5 K in the core to T = 266.8 K in the peripheral regions. The relative proportions reveal the predominance (52%) of epithermal neutrons in the reactor core, followed by thermal (29%) and fast (19%) neutrons. The moderation process proves to be effective, considering that 100% of the neutrons fall within the thermal spectrum near the reflector tank wall. The fits using Maxwellian distributions reveal another effect of neutron moderation: a shift in the peak wavelength from 1.07 Å to 1.19 Å. The same effect is observed along the tube that supplies a neutron beam to the RMB imaging instrument, with $\lambda_{peak}$ = 1.18 Å increasing to $\lambda_{peak}$ = 1.21 Å.

## 1. INTRODUCTION

Understanding neutron transport in research nuclear reactors is fundamental to enhance applications in material irradiation, radioisotope production, and neutron beam extraction [1]. The essential safety functions of a nuclear reactor—namely, the shutdown mechanism to maintain a subcritical state, efficient fuel cooling, and the containment of radioactive products—are critical for safeguarding personnel, the environment, and the facility under all operational conditions and potential faults. The implementation of a graded approach [2] enables these safety systems to be appropriately sized and applied based on the significance and associated risks of each function. This methodology optimizes resources by imposing the most stringent controls on areas and components with the highest potential impact, while proportionate requirements are established for elements deemed to be of lower risk. Consequently, this approach ensures an effective balance between safety, complexity, and cost.

Still assuming, for simplicity, a homogeneous reactor and adopting only a single neutron energy group (average) with a constant cross-section with respect to energy, we can express the reactor power as $P = E_f N_f \sigma_f \phi V$, where $E_f$ is the average energy, $N_f$ is the atomic concentration, $\sigma_f$ is th neutron reaction cross-section, $\phi$ is the neutron flux, and V is the core volume. Unlike power reactors, research reactors aim to maximize neutron flux to meet diverse objectives. Therefore,

$$\phi = \frac{P}{E_f N_f \sigma_f V} = \frac{\rho}{E_f N_f \sigma_f}, \qquad (1)$$

where $\rho \overset{\text{def}}{=} P/V$ is the reactor power density, most important to produccess high neutron flux. Achieving a high thermal neutron flux requires elevated reactor power, but power remains a primary limiting constraint. The compact core design allows for increased neutron flux without increasing the power level. This observation is made even clearer by the following equation presented by Difilippo and collaborators [3]

$$\phi_{th} \propto \frac{P}{A} = \frac{P^{2/3}}{V} = P^{1/3}\rho^{2/3}, \quad (2)$$

where A is the core area. High power density requires elevated flow velocities to ensure proper fuel cooling, which makes the structural integrity of the fuel assembly even more critical. To preserve this integrity, the flow velocity must remain below a threshold known as the critical flow velocity.

In multipurpose research reactors, the neutron flux distribution is non-uniform and undergoes dynamic changes during transient events, such as rapid power variations. Under these conditions, the typically linear relationship between flux and power [4] becomes complex, directly affecting local power density. Consequently, a rigorous spatial and temporal assessment is required to prevent overheating (hot spots) and damage to experimental targets.

In contrast to commercial power reactors, where power density is typically quite homogeneous and designed for high levels of continuous energy generation, research reactors often exhibit very high power densities in specific core regions, such as in the vicinity of irradiation channels. This is because these reactors are optimized to maximize neutron flux at specific points, aiming for efficiency in radioisotope production or the execution of experiments. For this reason, thermal management at these locations is critical, as the high power density can cause localized temperature increases, demanding efficient cooling systems and rigorous monitoring to prevent damage to the fuel elements or the material being irradiated.

Brazil is building its fifth nuclear reactor, called the Brazilian Multipurpose Reactor (RMB), in the city of Iperó, São Paulo (Figure 1). RMB is a 30 MW open-pool reactor with operations expected to begin by 2032. Its primary objective is to ensure national self-sufficiency in radiopharmaceuticals, particularly Mo-99, while also supporting nuclear fuel testing, advanced material irradiation, and neutron beam research [5]. The reactor project has a compact 5×5 core containing 23 $U_3Si_2$-Al fuel elements (enriched to 19.75%), designed to achieve a high neutron flux of $4 \times 10^{14}$ n·cm$^{-2}$·s$^{-1}$.

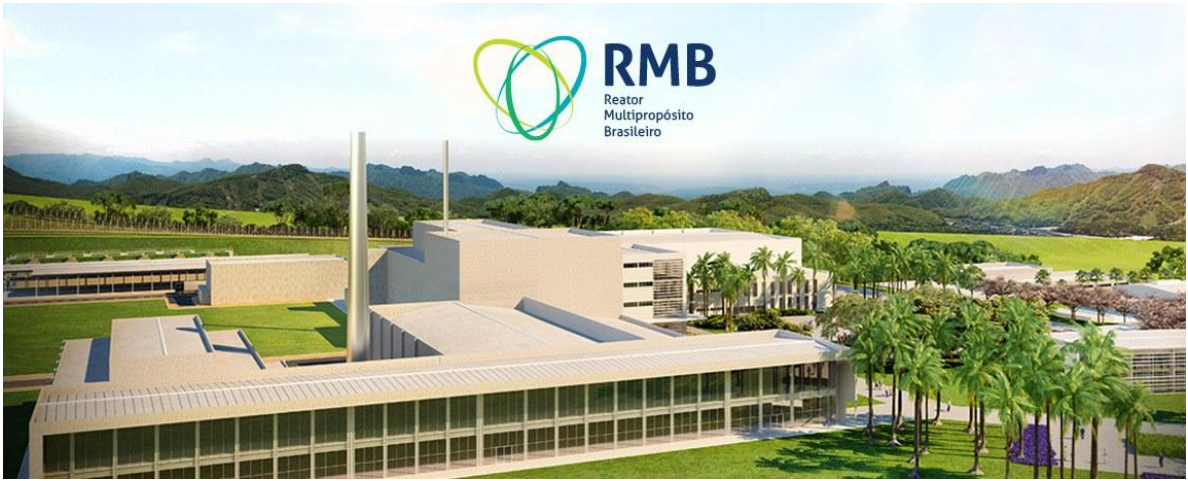

*Figure 1. Artistic view of the RMB reactor building.*

The objective of this study is to describe the neutron transport inside the RMB reflector tank through Monte Carlo simulations. For this task, we employed the MCNP6 (Monte Carlo N Particle) code [6] to model the reactor's reflector tank, which contains the fuel elements that compose the core, the beryllium and heavy water reflectors, the radionuclide irradiation positions, and the neutron beam guides. To capture the neutron distribution at specific points within the reflector tank, we used small spheres to record the neutron flux via the MCNP F2 tally card. In this work, we adopted the following energy ranges (E) which classify the neutrons as thermal (E < 0.65 eV), epithermal (0.65 eV < E < 0.81 MeV), and fast (E > 0.81 MeV).

## 2. MCNP6 MODEL

The simulations were performed on the Coaraci Supercomputer (Fapesp grant #2019/17874-0) at the University of Campinas, Brazil. Detailed informations on the construction of the reflector tank model is provided in reference [4,7]. The MCNP6 simulations used a *kcode* card with parameters 10⁷-1-50-1050, totaling 1050 cycles. The first 50 cycles were discarded to allow source convergence, and the neutron distributions were collected at the points indicated in Figure 2.

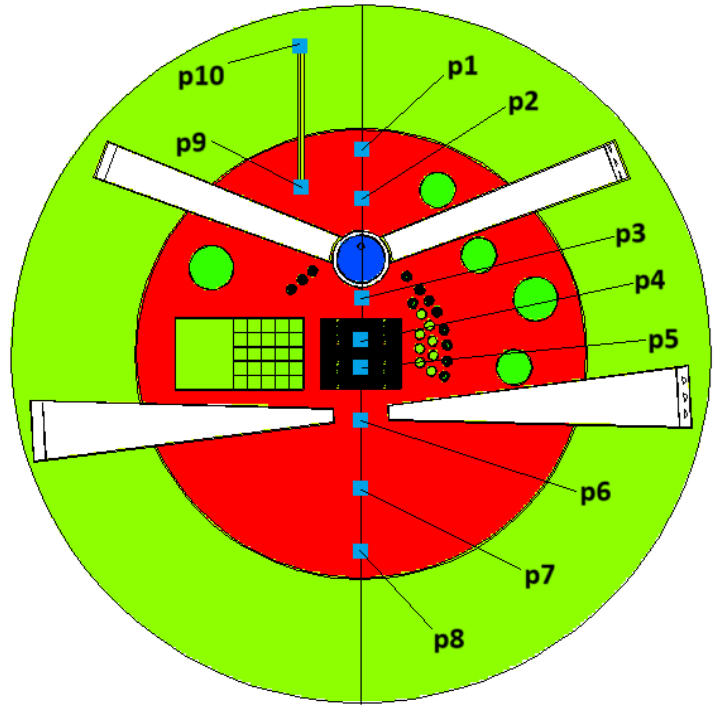

*Figure 2. Reflector tank model in MCNP6 code.*

According to transport theory [1], the thermal neutron spectrum can be approximately described by the function φ(λ), defined as

$$\phi(\lambda) = \sum_i^N \frac{\eta_i}{\lambda^5 T_i^2} \exp\left(-\frac{\gamma}{T_i \lambda^2}\right), \qquad (3)$$

where $\eta_i (n\ cm^{-2}s^{-1}K^2\text{Å}^5)$ and $\gamma(\text{Å}^2K)$ are constants, $T_i(K)$ is the medium temperature, N is the number of Maxwellian curves, and $\lambda$ is the neutron wavelength. The outputs obtained from the MCNP6 calculations allow us to determine the fitted parameters for the neutron flux at the different points.

With the data at hand, we are able to analyze the neutron beam thermalization process along the radial distance with respect to the reactor center, taking into account the presence of irradiation devices. In this manner, the percentage of each neutron spectrum range can be quantified, providing essential data for linking deposited power in specific regions.

## 3. RESULTS AND DISCUSSIONS

The relative proportions of thermal, epithermal, and fast neutrons at each of the points highlighted in Figure 2 are presented in Table I. We can observe that the data describe the neutron moderation process inside the RMB pool with good quality; that is, the fission reaction provides a spectrum of thermal, epithermal, and fast neutrons that undergoes thermalization. The population of epithermal and fast neutrons gradually decreases as the distance between the reactor core and the analyzed point increases.

*Table I. Relative proportion of thermal, epithermal, and fast neutrons at each analyzed point.*

| Point | Thermal (%) | Epithermal (%) | Fast (%) |
|-------|-------------|----------------|----------|
| p1 | 100 | 0 | 0 |
| p2 | 99.7 | 0.3 | 0 |
| p3 | 62 | 35 | 3 |
| p4 | 28 | 52 | 20 |
| p5 | 29 | 52 | 19 |
| p6 | 92 | 8 | 0 |
| p7 | 99.9 | 0.1 | 0 |
| p8 | 100 | 0 | 0 |
| p9 | 99.99 | 0.01 | 0 |
| p10 | 100 | 0 | 0 |

The Maxwellian approximation of the neutron distribution aims to provide a functional description for energies below 0.65 eV. Table II presents the Maxwellian parameters for the analyzed points, as well as the $R^2$ factor, which indicates the quality of the fits performed. It can be observed that the neutron distributions farther from the reactor core exhibit the best fits, resulting from the dominance of already thermalized neutrons.

*Table II. Results of the Maxwellian function parameters for N = 1.*

| Point | $T_1$ (K) | $\eta_1 (n\ cm^{-2}s^{-1}K^2\text{Å}^5)$ | $\gamma\left(\text{Å}^2 K\right)$ | $R^2$ | $\lambda_{peak}$ (Å) |
|-------|-----------|------------------------------------------|-----------------------------------|-------|----------------------|
| p1 | 266.8 | $1.4996 \times 10^{19}$ | 949 | 0.9715 | 1.19 |
| p2 | 266.3 | $3.2969 \times 10^{19}$ | 949 | 0.9658 | 1.19 |
| p3 | 290.5 | $1.2983 \times 10^{20}$ | 949 | 0.9669 | 1.14 |
| p4 | 331.1 | $4.5771 \times 10^{19}$ | 949 | 0.8948 | 1.07 |
| p5 | 331.5 | $5.0081 \times 10^{19}$ | 949 | 0.9084 | 1.07 |
| p6 | 273.4 | $1.3463 \times 10^{20}$ | 949 | 0.9791 | 1.18 |
| p7 | 273.7 | $5.1614 \times 10^{19}$ | 949 | 0.9676 | 1.18 |
| p8 | 280.7 | $1.9138 \times 10^{19}$ | 949 | 0.9735 | 1.16 |
| p9 | 273.0 | $1.0248 \times 10^{19}$ | 949 | 0.9896 | 1.18 |
| p10 | 258.7 | $2.4345 \times 10^{17}$ | 949 | 0.9837 | 1.21 |

We can observe a shift in the peak wavelength of the neutron distribution, an effect already expected due to moderation by light water. Three particular points stand out (p5, p9, and p10), for which we present the fitted curves in Figures 3–5, respectively. Figure 3 illustrates the neutron distribution at position p5, that is, in the reactor core. Figures 4–5 show the fitted curves at the entrance (p9) and exit (p10) of the neutron guide tube for the RMB imaging instrument [7].

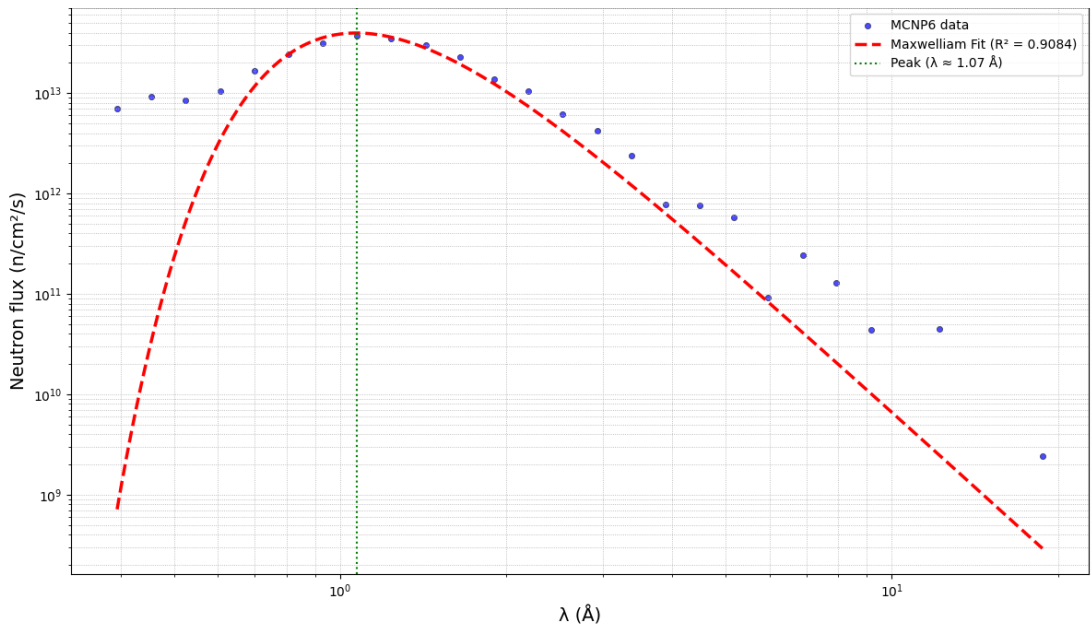

*Figure 3. Measured neutron flux distribution at p5.*

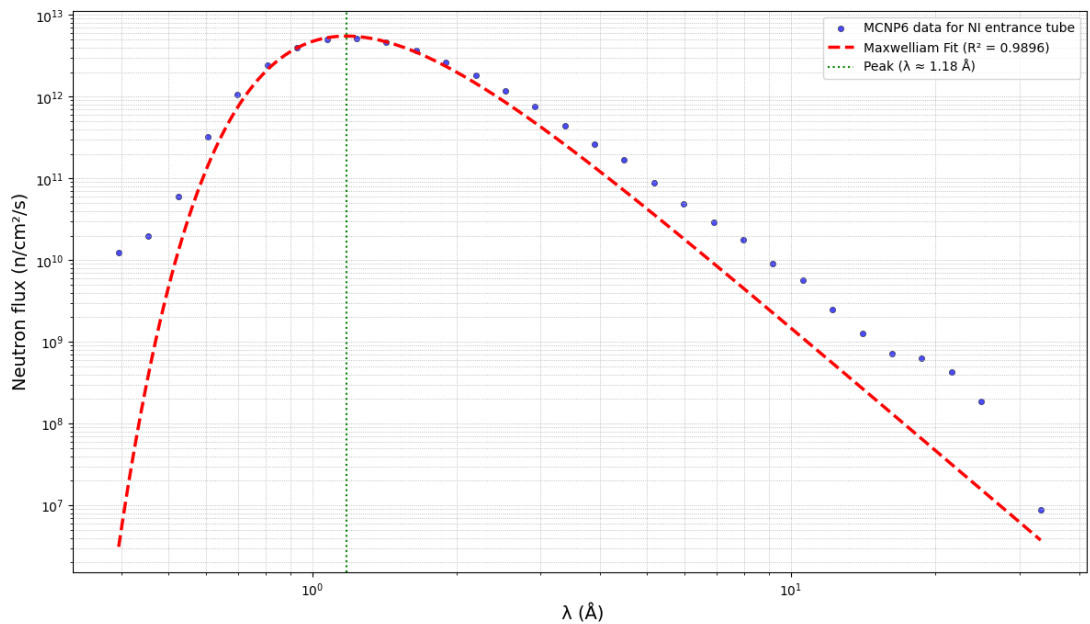

*Figure 4. Measured neutron flux distribution at p9.*

## 4. CONCLUSIONS

The neutron moderation process inside the RMB reflector tank was well described by the energy spectra at selected points, presented in Tables I and II. The relative proportions reveal the predominance (52%) of epithermal neutrons in the vicinity of the reactor core, followed by thermal (29%) and fast (19%) neutrons. A gradual decrease in temperature is observed, from T = 331.5 K at point p5 to T = 266.8 K at point p1, indicating the neutron moderation process along the radial dimension. The results of this article allow us to describe the behavior of neutron transport inside the RMB reflector tank, providing fundamental inputs for the operation and optimization of this new multipurpose reactor's objectives.

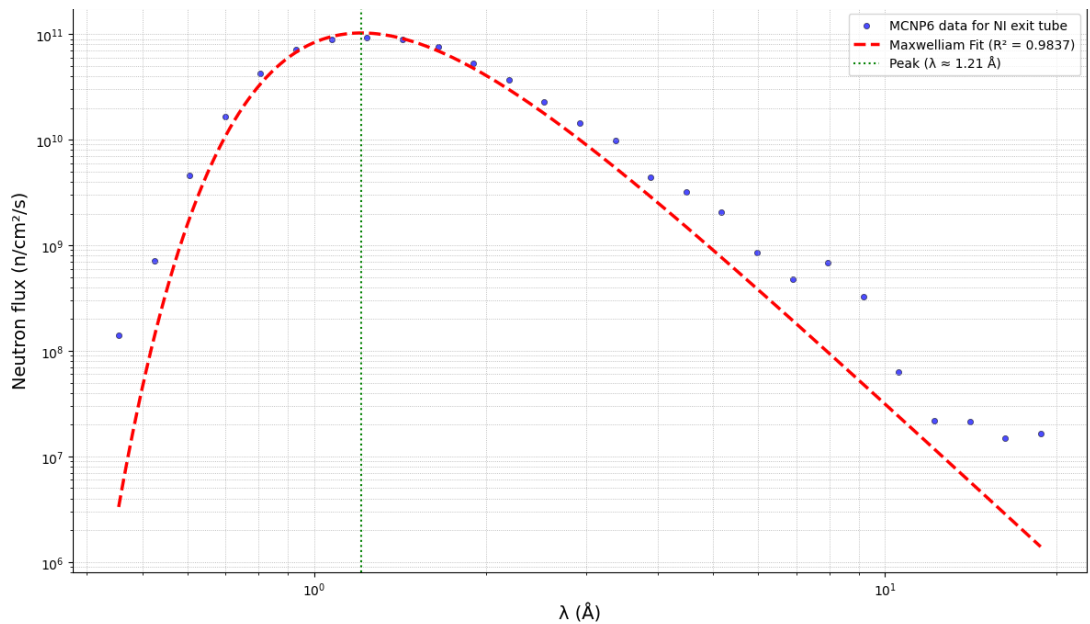

*Figure 5. Measured neutron flux distribution at p10.*

## ACKNOWLEDGMENTS

We thank the Coaraci Supercomputer for computer time (Fapesp grant #2019/17874-0) and the Center for Computing in Engineering and Sciences at Unicamp (Fapesp grant #2013/08293-7). C.G.S. Santos and I.S.R. Júnior thank CNEN/Fundação PATRIA for the research scholarships 006/2025 and 007/2025, respectively, under the FINEP agreement 01.24.0373.00 (RMB280).